\documentstyle[aps,prl,twocolumn]{revtex}
\input epsf
\begin{document}
\draft
\title{Brillouin propagation modes in optical lattices: 
       interpretation in terms of nonconventional stochastic resonance}

\author{L. Sanchez-Palencia, F.-R. Carminati, M. Schiavoni, 
        F. Renzoni and G. Grynberg}

\address{Laboratoire Kastler-Brossel, D\'epartement de Physique de l'Ecole
Normale Sup\'erieure, 24, rue Lhomond, 75231, Paris Cedex 05,
France}

\date{\today{}}

\maketitle
\begin{abstract}
We report the first direct observation of Brillouin-like 
propagation modes in a dissipative periodic optical lattice. This 
has been done by observing, in both theoretical and experimental work,
a resonant behaviour of the spatial diffusion coefficient in the direction 
corresponding to the propagation mode with the phase velocity of the 
moving intensity modulation used to excite these propagation modes. 
Furthermore, we show theoretically that the amplitude of the Brillouin mode
is a nonmonotonic function of the strength of the noise corresponding
to the optical pumping, and discuss this behaviour in terms of nonconventional
stochastic resonance.

\end{abstract}
\pacs{05.45.-a, 42.65.Es, 32.80.Pj}

The last decade has witnessed dramatic progress in laser
cooling techniques and nowadays in several laboratories around the
world atoms are routinely trapped and cooled at very low
temperatures and high densities \cite{metcalf}. Most of the
current efforts within the cold atoms community are directed to
reaching the regime of quantum degeneracy in both bosonic and
fermionic samples, in order to investigate the properties of the
{\it quantum gases} thus obtained, and realizing an atom laser, the
matter wave analog of the laser. Cold atomic samples also
constitute an ideal system for the study of complex nonlinear
phenomena. This turns out to be especially true if the cold atoms
are ordered by the light fields in periodic structures, 
so-called optical lattices \cite{jessen,robi}. These are obtained by
the interference of two or more laser fields: the light imposes its
order on the matter via the dipole force \cite{grimm}, creating a 
periodic structure of atoms.

Among the most significant studies of nonlinear dynamics in
optical lattices, we recall here the observation of mechanical
bistability in a strongly driven dissipative optical lattice
\cite{grynberg00} and the realization of the kicked rotor and
corresponding detection of chaotic motion in a far detuned lattice
\cite{raizen}. Furthermore the macroscopic transport of atoms in an 
asymmetric optical lattice without the application of external forces 
has been observed \cite{robi99}. This corresponds to the realization of
an {\it optical motor}, i.e. a ratchet for cold atoms, a well-controlable 
model system for the molecular combustion motor \cite{bartu}.
Brillouin-like propagation modes of atoms in a dissipative 
optical lattice have also been theoretically studied to 
explain the nonlinear optical properties of optical lattices
\cite{courtois96,brillo}. In this Letter, we report on the first 
direct observation of these modes. Furthermore we will discuss
the propagation mechanism associated with these modes, completely 
different from the one encountered in dense fluids or solid 
media. Indeed in dilute optical lattices the interaction between 
the different atoms is completely negligible, therefore the 
mechanism for the propagation of atoms cannot be ascribed to 
any sound-wave-like mechanism. While a sound wave corresponds to a 
propagating density wave without a net transport of atoms, the 
Brillouin-like resonances analyzed in this work consist of a net 
motion of the atoms. In fact, the propagation of atoms 
through the lattice are determined here by the interaction with the
light. The light fields determine both the potential wells where the
atoms can oscillate, and whose vibrational frequency determines the 
velocity of the propagation modes, and the escape from the potential wells,
which allows the propagation of atoms. We also show theoretically 
that the amplitude of the Brillouin mode is a nonmonotonic function of
the strength of the noise corresponding to the optical pumping, and discuss
this behaviour in terms of {\it nonconventional stochastic resonance}
\cite{gamma,dykman}.
%%%%%%%%%%%%%%%%%%%%%%%%%%%%%%%%%%%%%%%%%%%%%%%%%%%%%%%%%%%%%%%%%%%%%%%%%%%
\begin{figure}[ht]
\begin{center}
\mbox{\epsfxsize 3.in \epsfbox{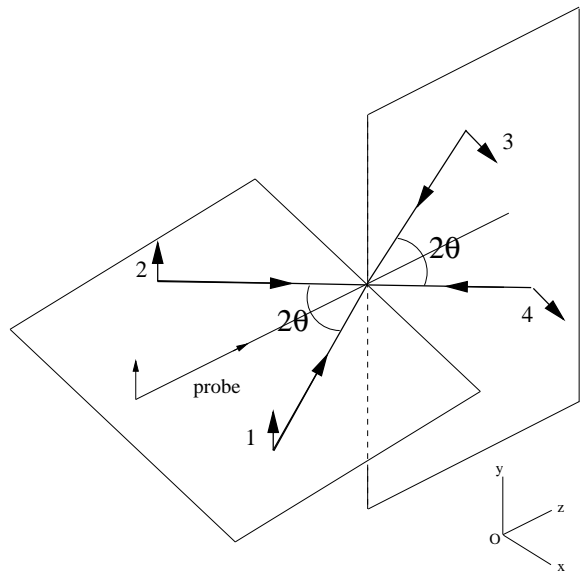}}
\end{center}
\caption{Sketch of the experimental setup.}
\label{fig1}
\end{figure}
%%%%%%%%%%%%%%%%%%%%%%%%%%%%%%%%%%%%%%%%%%%%%%%%%%%%%%%%%%%%%%%%%%%%%%%%%%%
We consider a three dimensional (3D) lin$\perp$lin near resonant optical 
lattice, as in previous work \cite{epj1}. The periodic structure is 
determined by the interference of four linearly polarized laser beams, 
arranged as in Fig. \ref{fig1}.  This arrangement results in a periodic
modulation of the light polarization and light intensity, which produces 
a periodic modulation of the light shifts of the different ground states of
the atoms (optical potentials) \cite{robi}. The optical
pumping between the different atomic ground states combined with
the spatial modulation of the light shifts leads then to the
cooling of the atoms \cite{dalibard89} and to their localization
\cite{castin91} at the minima of the optical potentials, thus
producing a periodic array of atoms. 

After the cooling phase, characterized by a significant reduction
of the atomic kinetic temperature and by the creation of a periodic spatial
order, the atoms keep interacting with the light undergoing optical pumping 
cycles. The optical pumping may transfer an atom from a potential well to a 
neighbouring one, giving rise to a variety of transport phenomena 
\cite{epj1,epj2}. Among these, there are modes which correspond to 
the propagation of atoms through the optical lattice in a given direction. 
They consist of a sequence in which one half oscillation in a potential 
well is followed by an optical pumping process to the neighbouring well, 
and so on. One can estimate their velocity by $v_i=\lambda_i \Omega_i/(2\pi)$
where $\lambda_i$ is the lattice constant and $\Omega_i/(2\pi)$ the  
vibrational frequency in the $i-$direction \cite{brillo}. These modes 
were first identified through Monte-Carlo simulations in
Ref. \cite{brillo} and shown to produce resonance lines in the
nonlinear optical response of optical lattices. However up to now
no direct observation of these modes has been reported. This is
achieved in the present work by observing a resonant behaviour 
of the spatial diffusion coefficient in the direction corresponding
to the propagation mode with the phase velocity of the moving intensity 
modulation used to excite these propagation modes.

\begin{figure}[ht]
\begin{center}
\mbox{\epsfxsize 3.in \epsfbox{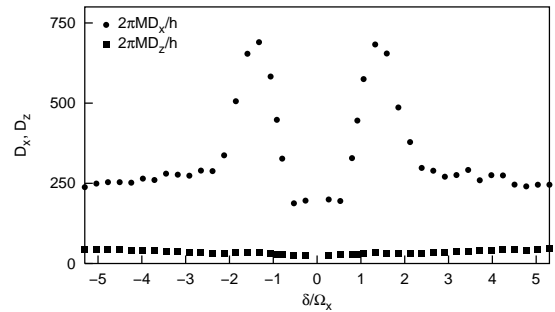}}
\end{center}
\caption{Numerical results for the spatial diffusion coefficients in the
$x$ and $z$ directions  as functions of the probe field detuning.
The lattice beam angle is $\theta=30^0$, the lattice detuning
$\Delta=-10\Gamma$ and the
light shift per beam $\Delta_0^{'}=-200 \omega_r$. Here $\Gamma$ and
$\omega_r$ are the width of the excited state and the atomic recoil
frequency, respectively. The amplitude of the probe beam is $0.4$ times
that of each lattice beam.}
\label{fig2}
\end{figure}

The modulation scheme for the excitation of the propagation modes
is completely analogous to the one used in previous investigations
of the nonlinear optical response of optical lattices \cite{courtois96}. 
An additional laser field linearly polarized along the $y$-axis is 
introduced with the $z$-axis as propagation direction.
This probe field interferes with the copropagating lattice beams, 
creating an intensity modulation. The interference pattern consists of
two propagating intensity waves moving with phase velocities 
$\vec{v}_j = \vec{n}_j \delta/|\Delta \vec{k}_j|$ ($j=1,2$) with 
$\vec{n}_j = \Delta \vec{k}_j/|\Delta \vec{k}_j|$, and
$\Delta \vec{k}_j =\vec{k}_j-\vec{k}_p$ the difference between the 
$j$-th lattice beam and the probe ($p$) wavevectors \cite{single}. Here 
$\delta=\omega_p-\omega_L$ is the detuning between the probe ($\omega_p$)
and the lattice ($\omega_L$) frequencies. According to the numerical 
simulations for the atomic trajectories presented in Ref. \cite{brillo},
for $\delta=\pm\Omega_x$, the propagation modes along $x$ are excited
by the driving field, with the atoms effectively dragged by the moving 
intensity modulation \cite{nb}. Intuitively, the dragging of atoms by 
the two propagating intensity modulations should result in an increase 
of the spatial diffusion coefficient $D_x$ in the $x$-direction. Therefore it 
should be possible to detect these Brillouin propagation modes by monitoring  
$D_x$ as a function of the detuning $\delta$. The propagation modes are
then revealed by a resonance in $D_x$ around $\delta=\pm\Omega_x$. We 
tested the validity of this reasoning with the help of semiclassical 
Monte-Carlo simulations \cite{mc}. Taking advantage of the symmetry between 
the $x$ and $y$ directions (see Fig. \ref{fig1}), we restricted the atomic 
dynamics in the $xOz$ plane. Our calculations are for a $J_g=1/2
\to J_e=3/2$ transition, as customary in numerical analysis of 
Sisyphus cooling, of an atom of mass $M$. We expect our 2D calculations 
to reproduce the dependencies of the different quantities associated with 
the real 3D atomic dynamics to within a scaling factor corresponding to the 
difference in dimensionality \cite{epj2}. In the numerical simulations, we 
monitored the variance of the atomic position distribution at a given
value of the probe field detuning. We verified that the spatial 
diffusion is normal, i.e. the atomic square displacements
$\langle\Delta x^2\rangle$ and $\langle\Delta z^2\rangle$ increase 
linearly with time. Accordingly, we derived the spatial diffusion
coefficients $D_x$ and $D_z$ by fitting the numerical data with
$ \langle\Delta x_i^2\rangle=2 D_{x_i} t$  $(x_i=x,z)~.$
Results for the spatial diffusion coefficients as functions of the 
probe detuning $\delta$ are shown in Fig. \ref{fig2}. Two narrow 
resonances, centered approximately at $\delta=\pm\Omega_x$, appear 
clearly in the spectrum of the diffusion coefficient along the $x$-axis. 
In contrast, $D_z$ does not show any resonant behaviour with the driving 
field detuning. This demonstrates the validity of the detection scheme
based on the measurement of the diffusion coefficients.
%%%%%%%%%%%%%%%%%%%%%%%%%%%%%%%%%%%%%%%%%%%%%%%%%%%%%%%%%%%%%%%%%%%%%%%%%%
%%%%%%%%%%%%%%%%%%%%%%%%%%%%%%%%%%%%%%%%%%%%%%%%%%%%%%%%%%%%%%%%%%%%%%%%%%%%

In the experiment $^{85}$Rb atoms are cooled and trapped in a 
magneto-optical trap (MOT). The MOT laser beams and magnetic field 
are then suddenly turned off. Simultaneously the four lattice beams 
are turned on and after $10$ ms of thermalization of the atoms in the
lattice the probe laser field is introduced along the $z$-axis.
We studied the transport of atoms in the optical lattice by observing
the atomic cloud expansion with a Charge Coupled Device (CCD) camera.
The procedure to derive the diffusion coefficients has been described
in detail in previous work \cite{epj1}, and we recall here only the
basic idea.
For a given detuning of the probe field we took images of
the expanding cloud at different instants after the atoms have been
loaded into the optical lattice.
From the images of the atomic cloud we derived the atomic mean square
displacement along the $x$- and $z$-axes.
%%%%%%%%%%%%%%%%%%%%%%%%%%%%%%%%%%%%%%%%%%%%%%%%%%%%%%%%%%%%%%%%%%%%%%%%%%%%
\begin{figure}[ht]
\begin{center}
\mbox{\epsfxsize 3.5in \epsfbox{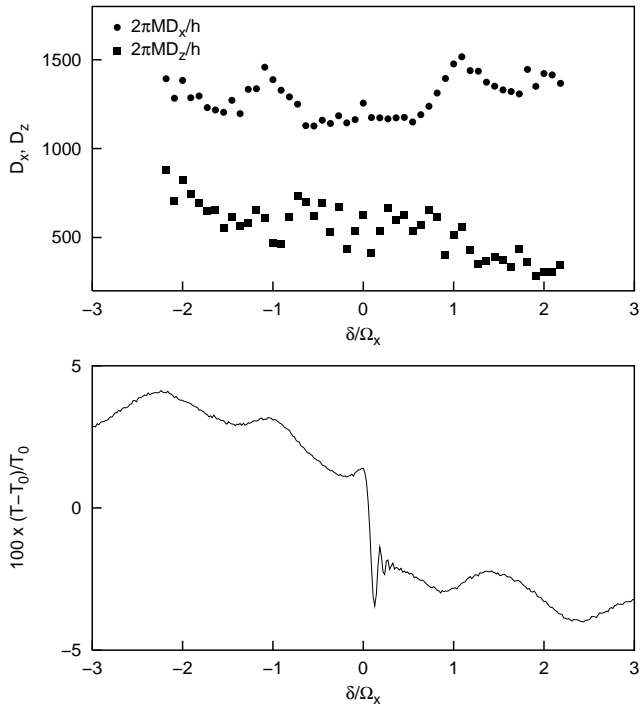}}
\end{center}
\caption{Experimental results for the spatial diffusion coefficients in
the $x$ and $z$ directions as functions of the probe field detuning.
The experimental parameters are: lattice detuning $\Delta/(2\pi) = -42$ MHz,
intensity per lattice beam $I_L = 3.5 $ mW/cm$^2$, lattice angle
$\theta=30^0$. These parameters correspond to a vibrational frequency
in the $x$- direction $\Omega_x/(2\pi) \simeq 55$ kHz. A probe transmission
spectrum is reported for comparison, $T$ and $T_0$ being the intensity
of the transmitted probe beam with and without the atomic cloud.
The two resonances at $\delta=\pm \Omega_x$ correspond to stimulated light
scattering by the Brillouin propagation modes. The two resonances at larger
detuning are Raman lines \protect\cite{robi} in the $z$-direction
($\delta=\pm \Omega_z$), which do not correspond to propagation modes.  For
the measurements of the diffusion coefficients, the probe beam intensity is
$I_p =  0.3$ mW/cm$^2$; in the transmission spectrum $I_p =  0.1$ mW/cm$^2$.}
\label{fig3}
\end{figure}
%%%%%%%%%%%%%%%%%%%%%%%%%%%%%%%%%%%%%%%%%%%%%%%%%%%%%%%%%%%%%%%%%%%%%%%%%%%%%
We verified that the cloud expansion corresponds to normal diffusion
and derived the diffusion coefficients $D_x$ and $D_z$. The procedure 
has been repeated for several different values of the detuning 
$\delta$ of the probe field. Results for $D_x$ and $D_z$ as functions
of $\delta$ are shown in Fig. \ref{fig3}. The probe transmission spectrum
is also reported to allow the comparison of the position of the resonances
in the spectrum of the diffusion coefficients and in that of the 
probe transmission.  We observe two narrow resonances in the diffusion 
coefficient along the $x$-axis centered at $\delta=\pm\Omega_x$. In 
contrast, no resonant behaviour of $D_z$ with $\delta$ is observed. This 
is in agreement with the numerical simulations and constitutes the first 
direct observation of Brillouin-like propagation modes in an optical lattice.

\begin{figure}[ht]
\begin{center}
\mbox{\epsfxsize 3.5in \epsfbox{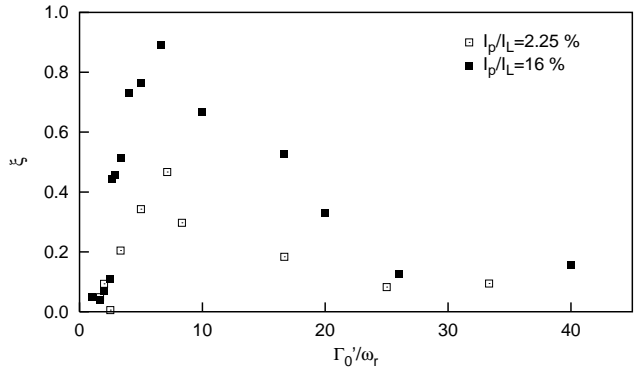}}
\end{center}
\caption{
Numerical results for the enhancement factor $\xi$ as a function of the
optical pumping rate, for a given depth of the optical potential wells.
Parameters for the calculations are: $\Delta_0' = -50 \omega_r$ and
$\theta=30^0$. The two data sets correspond to different intensities of
the probe beam. For comparison, we recall that in the experiment (Fig.
\protect\ref{fig3}) $\Delta_0' \simeq -60 \omega_r$ and
$\Gamma^{'}_0\simeq 8.5 \omega_r$.}
\label{fig4}
\end{figure}

We turn now to the analysis of the mechanism behind these propagation 
modes. Brillouin-like propagation modes have been widely studied in 
condensed matter and dense fluids \cite{books}. However in the present case 
the mechanism associated with these modes is clearly of a different nature, 
as in dilute optical lattices the interaction between atoms is negligible
and therefore sound-wave-like propagation modes cannot be supported.
On the contrary, in a dilute optical lattice the propagation of the 
atoms is determined by the synchronization of the oscillation within 
a potential well with the optical pumping from a well to a neighbouring 
one, as first identified in the numerical analysis of Ref. \cite{brillo}.
This dynamics can be interpreted in terms of noise-induced resonances:
the probe field induces a large scale moving modulation of the periodic 
potential of the four-beam optical lattice, with the optical pumping 
constituting the noise source which allows transfer from a well to a 
neighbouring one.  It is then natural to investigate the dependence of the 
amplitude of the Brillouin mode on the strength of the noise, i.e. on the
optical pumping rate.
We studied, via semiclassical Monte-Carlo simulations, the atomic 
cloud expansion for a given depth of the potential well at different values 
of the optical pumping rate $\Gamma_0'$, proportional to the rate
$\Gamma^{'}_{esc}$ of escape from the well \cite{rate}.  
This has been done by varying the lattice intensity $I$ and detuning 
$\Delta$ so as to keep the depth of the potential wells $U_0\propto I/\Delta$
constant while varying $\Gamma'_0 \propto I/\Delta^2$.
The diffusion coefficient in the $x$-direction
has been calculated both for a probe field at resonance ($|\delta|=\Omega_x$)
and for a probe field far off-resonance ($|\delta|\gg\Omega_x$). The two
diffusion coefficients will be indicated by $D_x$ and $D_x^0$ respectively.
To characterize quantitatively the response of the atomic system to a noise
strength variation, we introduce the enhancement factor $\xi$ defined as
\begin{equation}
\xi = \frac{D_x - D_x^0}{D_x^0}~.
\label{xi}
\end{equation}
Numerical results for the enhancement factor $\xi$ as a function
of the optical pumping rate at a given value of the potential well depth
(i.e. for fixed light shift per beam $\Delta_0'$) are shown in Fig.
\ref{fig4}. At small pumping rates, $\xi$ increases abruptly with $\Gamma_0'$; 
then a maximum is reached, corresponding to the synchronization of the 
oscillation of the atoms within a well with the escape from a well to the 
neighbouring one; finally at larger pumping rates this synchronization is 
lost and $\xi$ decreases. This dependence recalls  the typical behaviour 
of stochastic resonance \cite{gamma}, with the noise enhancing the response
of the atomic system to the weak moving modulation. It should be noted that 
the system analyzed here has one important peculiarity with respect to the 
model usually considered in the analysis of stochastic resonance.  
Stochastic resonance is in general understood as the noise-induced 
enhancement of a weak periodic signal with a frequency much smaller than 
the intrawell relaxation frequency within a single metastable state. In 
contrast, in the present case, the noise synchronizes precisely with the 
intrawell motion of the atoms. This corresponds to a nonconventional
stochastic resonance scenario \cite{dykman}.

%%%%%%%%%%%%%%%%%%%%%%%%%%%%%%%%%%%%%%%%%%%%%%%%%%%%%%%%%%%%%%%%%%%%%%%%%%%%

In summary, in this Letter we introduced a scheme for the detection of
Brillouin propagation modes in optical lattices and we 
reported on their direct observation. Furthermore, we studied 
via Monte-Carlo simulations the amplitude of the Brillouin mode,
as characterized by an increase of the diffusion coefficients due to the
presence of the probe field, as a function of the rate of escape from
the potential wells. 
The Brillouin modes examined in this work differ from their counterparts
in solid state or dense fluids as they are sustained by a medium of
{\it noninteracting} particles. From our analysis it turns out that in 
the presence of noise the Brownian motion of a system of particles in a 
periodic potential can be turned in a motion at a well-defined velocity by 
the application of a weak moving modulation. This represents a quite unusual 
situation in statistical physics and may constitute a model for many
biological phenomena, like the transmission of weak signals in neuronal 
systems \cite{brain}.

We thank Franck Lalo\"e for his continued interest in our work
and David Lucas for comments on the manuscript.
This work was supported by CNRS, the European Commission 
(TMR network "Quantum Structures", contract FMRX-CT96-0077) and 
by R\'egion Ile de France under contract E.1220.
Laboratoire Kastler Brossel is an "unit\'e mixte de recherche de
l'Ecole Normale Sup\'erieure et de l'Universit\'e Pierre et Marie
Curie associ\'ee au Centre National de la Recherche Scientifique
(CNRS)".


\begin{thebibliography}{99}
\bibitem{single}
To induce a propagating wave in a single direction a slightly more complicated
setup would be needed, with two additional laser beams,
independent of the lattice beams, creating the moving intensity modulation.
\bibitem{nb}
We recall that in this geometry the Raman excitation of the $\Omega_x$ 
vibration mode is forbidden \protect\cite{brillo}.
\end{thebibliography}
\end{document}